\documentclass[useAMS,usenatbib,usechicago]{mn2e}

\usepackage{graphicx}

\def\aca{Acta Astr.}%
\def\apj{ApJ}%
\def\aap{A\&A}%
\def\aaps{A\&AS}%
\def\mnras{MNRAS}%
\def\PST{Pozna\'n Spectroscopic Telescope}%

\title[V440 Per: the overtone Cepheid]{V440 Per: the longest period overtone Cepheid}

\author[Baranowski et al.]{R.~Baranowski$^{1}$\thanks{Based on observations from \PST},
        R.~Smolec$^{2}$, W.~Dimitrov$^{1}$, T.~Kwiatkowski$^{1}$, \newauthor
        A.~Schwarzenberg-Czerny$^{1,2}$\thanks{e-mail alex{@}camk.edu.pl}, P.~Bartczak$^{1}$,
        M.~Fagas$^{1}$, W.~Borczyk$^{1}$, \newauthor K.~Kami\'nski$^{1}$, P.~Moskalik$^{2}$,
        R.~Ratajczak$^{1}$ and A.~Ro\.{z}ek$^{1}$ \\
        $^{1}$ Astronomical Observatory of Adam Mickiewicz University, ul. S{\l}oneczna 36, PL 60-286 Pozna\'n,
         Poland, \\
        $^{2}$ Copernicus Astronomical Centre, ul. Bartycka 18, PL 00-716 Warsaw, Poland
}

\date{Accepted .......
      Received .......;
      in original form ........}

\pagerange{\pageref{firstpage}--\pageref{lastpage}} \pubyear{2009}

\begin{document}

\maketitle

\label{firstpage}

   \begin{abstract}
V440~Per is a Population~I Cepheid with the period of
$7.57$\thinspace day and low amplitude, almost sinusoidal light and
radial velocity curves. With no reliable data on the 1st harmonic,
its pulsation mode identification remained controversial. We
obtained a radial velocity curve of V440~Per with our new high
precision and high throughput {\PST}. Our data reach the accuracy of
130\thinspace m/s per individual measurement and yield a secure
detection of the 1st harmonic with the amplitude of $A_2= 140\pm
15$\thinspace m/s. The velocity Fourier phase $\phi_{21}$ of V440~Per is
inconsistent at the 7.25$\sigma$ level with those of the fundamental mode
Cepheids, implying that the star must be an overtone Cepheid, as
originally proposed by \citet{kie99}. Thus, V440~Per becomes the
longest period Cepheid with the securely established overtone
pulsations. We show, that the convective nonlinear pulsation hydrocode
can reproduce the Fourier parameters of V440~Per very well. Requirement
to match the observed properties of V440~Per constrains free parameters
of the dynamical convection model used in the pulsation
calculations, in particular the radiative losses parameter.
   \end{abstract}



\section{Introduction} \label{s0}

The Population I {\em sinusoidal} or s-Cepheids are a small group of
the Cepheids pulsating in the first radial overtone. In the Galaxy,
where individual Cepheid distances are usually not accurately known,
the s-Cepheids are discriminated from the fundamental mode pulsators
with the Fourier decomposition of their light curves \citep{APR90}.
The method works well only for the variables with periods below
5\thinspace days. Fortunately, it can be extended to longer periods
with the help of the radial velocity curves, as was shown by
\citet{kie99}. Studies of the longer period overtone Cepheids are of the great
interest. The velocity Fourier parameters of the s-Cepheids display a very
characteristic progression with period, which is attributed to the
2:1 resonance at $P\! =\! 4.2\! -\! 4.6$\thinspace day between the
first and fourth overtones \citep{kie99, FEU00}. Unfortunately,
any detailed modeling of this progression and pinpointing of the resonance
position is hampered by a scarcity of the s-Cepheids with $P >
5.5$\thinspace day, with MY~Pup being the only secure
identification. One more candidate, V440~Per ($P = 7.57$\thinspace
day), has been identified by \citet{kie99}. They noted, that the
velocity Fourier phase $\phi_{21}$ places this Cepheid away from the
fundamental mode sequence and possibly onto the first overtone
sequence. On this basis, \citet{kie99} proposed that V440~Per is an
overtone pulsator. This hypothesis was further supported by
determination of the phase-lag between the light curve and the
radial velocity curve, $\Delta\Phi_1=\phi_1^{V_r}-\phi_1^{mag}$,
which placed V440~Per away from the fundamental sequence as well
\citep{Og00}. However, V440~Per is a small amplitude, nearly
sinusoidal variable. Even the best then available radial velocity
data \citep{BB82} yielded large errors of the 1st harmonics Fourier
parameters. The measurement error of the phase-lag $\Delta\Phi_1$
was large, too. This, and inconsistency with their hydrodynamic
pulsation models led \citet{SZ07} to dispute the mode identification of
V440~Per. They noted that the membership of this Cepheid in the
fundamental mode sequence could not be rejected at the $3\sigma$
confidence level. They argued that V440~Per is not an overtone
pulsator, but rather a fundamental mode Cepheid of a very low
amplitude.

In the present paper we report results of an extensive campaign of
the observations of V440~Per with the {\PST}, lasting almost a year and
aiming at obtaining high quality data suitable for the detailed
diagnostic and comparison with the nonlinear models. In
Sect.\thinspace\ref{s1} we describe our instrument and observations.
In Sect.\thinspace\ref{s2} we discuss quality of our radial
velocities and derive Fourier parameters of V440~Per. Comparison
with Fourier parameters of other Galactic Cepheids and
identification of the pulsation mode of V440~Per is discussed in
Sect.\thinspace\ref{s3}. In Sect.\thinspace\ref{RSSEC1} we perform
detailed comparison of V440~Per with the nonlinear overtone Cepheid
models. Our conclusions are summarized in Sect.\thinspace\ref{s4}.

\section{Observations and data reduction} \label{s1}

Our observations were obtained with the new {\PST} (PST) of Adam
Mickiewicz University. Its full description is not published yet,
hence we devote some space here to instrument description and data
quality evaluation.

\subsection{\PST}

PST is located in Poland at Borowiec station, 20\thinspace km south
from Pozna\'n city, at a meagre elevation of 123\thinspace m above see
level. The PST consists of parallel twin 0.4\thinspace m Newton
telescopes of the f-ratio 4.5, fixed on a single parallactic fork mount.
An acquisition box at Newton focus of each telescope holds the tip of the
fiber feeding our spectrograph, the Thor/Argon calibration lamp and
the autoguider camera SBIG ST-7.

The telescope feeds via a fiber an Echelle spectrograph, a clone of
MUSICOS design \citep{bau92}, red arm
only. Our spectra are recorded with the low noise Andor DZ436 camera
fitted with $2k\!\times\! 2k$ E2V 42-40 back illuminated CCD chip,
cooled with Peltier cells. About 60 orders are recorded, covering a
spectral range of 4480\thinspace --\thinspace 9250\thinspace {\AA}
at the inverse resolution of $\lambda/\delta\lambda = 35000$. The
spectrograph is located in a thermo-isolated enclosure in the
telescope dome. The sliding-roof dome, the telescope and its spectrograph all
operate under full computer control. A full description of our instrumentation will be
published elsewhere. Our system is operated interactively from the terminal. Such an operation mode
does not compromise our data quality but requires excessive work-load.
A substantial software effort is needed to achieve the fully robotic operation.

\begin{table}
\begin{center}
\caption{PST radial velocities of V440~Per.\newline Full table in the electronic form only.
         \label{t1}}
\begin{tabular}{r@{\hspace{0.04\textwidth}}rr}
\multicolumn{1}{c}{MJD\hspace{0.04\textwidth}}
                & \multicolumn{1}{c}{$\Delta v_{r}^{a)}$} 
                & \multicolumn{1}{c}{$\Delta v_{r,corr}^{b)}$} \\
\hline
  54327.0490 &      -0.61 &      -0.52 \\
  54332.0517 &       3.04 &       3.14 \\
  54332.0683 &       2.96 &       3.06 \\
\multicolumn{3}{c}{$\cdots$} \\
  54649.0009 &       2.28 &       2.12 \\
  54650.9302 &       2.75 &       2.57 \\
  54650.9419 &       2.76 &       2.58 \\
\hline
\end{tabular}\\
\begin{minipage}[l]{0.35\textwidth}
 \footnotesize a) zero shifted by about $-26.3\pm0.2$ km/s\\
 \footnotesize b) corrected for instrumental effects (see text)
\end{minipage}
\end{center}
\end{table}

\begin{figure}
\includegraphics[height=0.49\textwidth, angle=270]{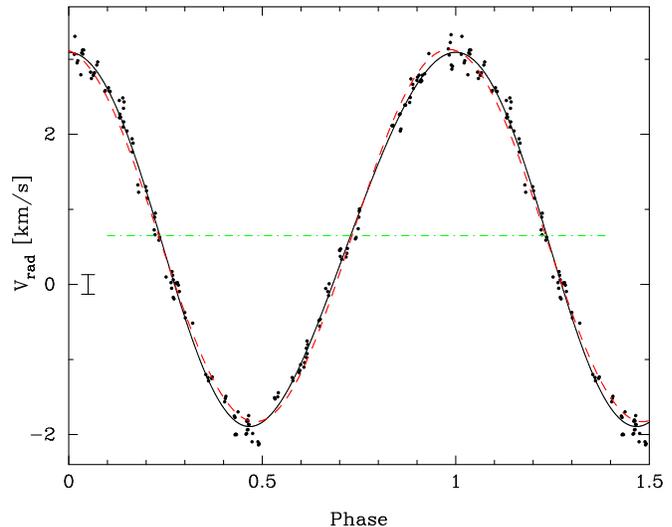}
\caption{Radial velocities of V440~Per {\em vs.} phase. First and
         second order Fourier fits are plotted with dashed and solid
         lines, respectively. The error bar indicates $\pm
         $\thinspace 1 standard deviation of residuals. The
         velocities are corrected for small instrumental effects
         (see text). Zero point is arbitrary.}
\label{ASC1}
\end{figure}

\begin{figure}
\includegraphics[height=0.47\textwidth, angle=270]{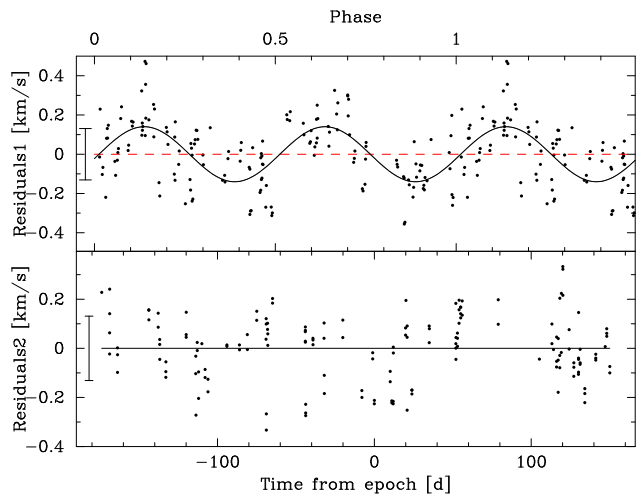}
\caption{Top\thinspace : residuals of the first order Fourier fit of
         V440~Per radial velocities (dashed line in
         Fig.\thinspace\ref{ASC1}) {\em vs.} phase. 1st harmonic is
         very small ($A_2= 140\pm 15$\thinspace m/s), yet clearly
         visible. Bottom\thinspace : residuals of the second order
         Fourier fit {\em vs.} time.}
\label{ASC2}
\end{figure}


\subsection{Data pipeline}

Routine CCD reductions up to spectra extraction, wavelength
calibration and velocity measurement are performed using IRAF
tasks combined into our reduction pipeline. Velocities are
measured by cross-correlation, using IRAF FXCOR task. The internal
error estimates from FXCOR, of order of few km/s, relate to the
line width and not to the actual measurement precision. They may
serve for weighting purposes, though. So far, we employed no
standard star calibrations and our velocities are measured solely
with respect to the Th/Ar lamp. For short span observations such a
primitive procedure still yields root-mean-square (RMS) residuals
in the 100\thinspace --\thinspace 200\thinspace m/s range, for
stars brighter than 11\thinspace mag. As demonstrated in
Sect.\thinspace\ref{s2}, for intensively observed, strictly
periodic stars, any long-term effects of a floating instrumental
zero-point may be removed with some assurance, yielding again RMS
residuals less than 150\thinspace m/s, over one year span of data.

\subsection{Observations}

In total 158 radial velocity measurements of V440~Per were
obtained with PST from 2007 August 15 till 2008 July 03. To reach
signal-to-noise ratio of 70 we exposed spectra for 10\thinspace
--\thinspace 15\thinspace min and up to several spectra were
obtained per night. The observed velocities are listed in
Table~\ref{t1} published in the electronic form. Phase-folded data
with fitted Fourier series (Sect.\thinspace\ref{s2}) are displayed
in Fig.\thinspace\ref{ASC1}. In Fig.\thinspace\ref{ASC2} we plot
residuals of the first order and the second order fits {\em vs.}
pulsation phase and {\em vs.} time. Inspection of the plots
demonstrates that our coverage was reasonably uniform, both in
time and in frequency. From the residuals we estimate standard
error of our individual measurements as 130\thinspace m/s.

\begin{table}
\begin{center}
\caption{Fourier parameters of V440~Per radial velocity curve
         \label{t2}}
\begin{tabular}{lllc}
Name         & Value     & Error  & Unit \\
\hline
 $T_0$       & 54498.783 & 0.008  & MJD  \\
 $P$         & 7.5721    & 0.0006 & d    \\
 $A_1$       & 2.480     & 0.015  & km/s \\
 $A_2$       & 0.140     & 0.015  & km/s \\
 $R_{21}$    & 0.056     & 0.006  & 1    \\
 $\phi_{21}$ & 2.759     & 0.117  & rad  \\
\hline
\end{tabular}
\end{center}
\end{table}

\section{Fourier parameters} \label{s2}

The plot of the phase-folded radial velocities reveals near
sinusoidal variations with peak-to-peak amplitude of 5\thinspace
km/s. Some data consistency checks are due prior to drawing any
final conclusions. Cepheid phases and frequencies are known to vary
slightly. Additionally, the instrument stability over 9\thinspace
months needs checking, too. A preliminary nonlinear least squares
fit of our data with the Fourier series of three harmonics terms
(third order fit) yielded no significant 2nd harmonics of the main
frequency nor the period derivative term. Our fitted frequency of
$0.13206\pm 0.00001$\thinspace c/d is less accurate, yet consistent
within errors with the frequency derived by combining our data with
the earlier measurements of \citet{BB82}, \citet{arr84} and
\citet{Gor92,Gor96,Gor98}. Comparison of the data shows, that the
weighted zero point shift of our velocities with respect to the
previous authors is $-26.3\pm 0.2$\thinspace km/s.

The RMS deviation from the Fourier fit of all our data was
164\thinspace m/s, well in excess of 135\thinspace m/s obtained for
the first part of the dataset. Worse, inspection of the residuals
plotted against time sometimes revealed non-gaussian, bi-modal
distribution. The origin of both effects seems to be instrumental.
To confirm that, we expanded our Fourier model by including two
linear terms proportional to the time interval from the mid-epoch
and to the hour angle of the star at the moment of observation. The
fitted values of these instrumental correction coefficients were
significant at 6$\sigma$ and 5$\sigma$ level, respectively. From the
overall covariance matrix we find maximum absolute value of the
correlation coefficients of 0.30, consistent with little
interference between different fitted terms. These instrumental
corrections reached up to $\pm 100$\thinspace m/s. At this stage, we
have no explanation for these corrections.

Our final second order Fourier fit, supplemented with the
instrumental correction terms, yielded RMS deviation of
130\thinspace m/s, consistent with that obtained from the short-span
observations. The values of the Fourier parameters of V440~Per
radial velocity curve are listed in Table~\ref{t2} (for exact
formulae defining Fourier parameters and their errors see
Appendix~\ref{sa}).

\section{Pulsation mode of V440~Per} \label{s3}

\begin{figure}
\includegraphics[height=0.47\textwidth, angle=270]{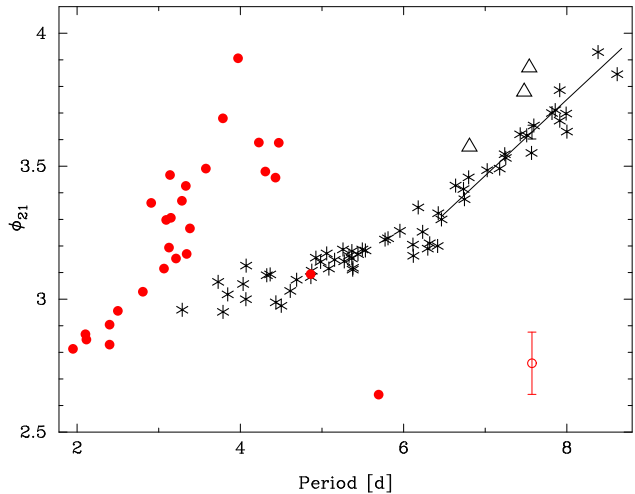}
\caption{Fourier phase $\phi_{21}$ {\em vs.} pulsation period for
         Cepheid radial velocity curves. Fundamental mode Cepheids
         displayed with asterisks, except for low amplitude ones
         ($A_1<10$\thinspace km/s) marked with open triangles.
         Overtone Cepheids displayed with filled circles.
         Observational data points taken from \citet{kie99} and
         \citet{MGS09}. Straight line indicates the best fit to the
         segment of the fundamental mode progression. V440~Per
         displayed with open circle.}
\label{ASC3}
\end{figure}


Pulsation mode of a Cepheid can be established by measuring Fourier
phase $\phi_{21}$ of its light curve \citep{APR90} or radial
velocity curve \citep{kie99}. This Fourier parameter does not depend
on the pulsation amplitude of the star and for each mode it follows a
different, {\em tightly defined} progression with the pulsation
period. For periods at which the two $\phi_{21}$ progressions are
well separated, secure mode identification can be achieved.

In Fig.\thinspace\ref{ASC3} we plot velocity $\phi_{21}$ of short
period Galactic Cepheids against their pulsation period $P$.
Fundamental mode pulsators and overtone pulsators are displayed
with different symbols. V440~Per is plotted with an open circle.
It is immediately obvious, that it is located far apart from the
fundamental mode progression. This notion can be put on a
quantitative basis. We selected a sample of 23 fundamental mode
Cepheids with periods $P$ in the range of $P_0\pm1.1$\thinspace
day, where $P_0=7.5721$\thinspace day is the period of V440~Per.
To this sample we fitted a straight line $\phi_{21}(P) =
a(P-P_0)+b$. With this procedure, we find that at the period of
V440~Per the expected $\phi_{21}$ of the fundamental mode Cepheid
is 3.628\thinspace rad, with the average scatter of individual
values of $\sigma_0=0.026$\thinspace rad. This estimate of the
intrinsic scatter is conservative, as the nominal $\phi_{21}$
measurement errors would account for at least half of it. The
$\phi_{21}$ value measured for V440~Per is 2.759\thinspace rad,
with an error of $\sigma_1=0.117$\thinspace rad. The probability
distribution of the $\phi_{21}$ offset, $\Delta$, is obtained by
convolution of two normal distributions, $N(0,\sigma_0)$ and
$N(\Delta,\sigma_1)$. It may be demonstrated, that the result is
another normal distribution
$N(\Delta,\sqrt{\sigma_0^2+\sigma_1^2})$, essentially by virtue of
the law of error combination. By substituting
$\Delta=3.628-2.759=0.869$\thinspace rad, we find that the observed
velocity $\phi_{21}$ of V440~Per deviates from the fundamental
mode sequence by 7.25$\sigma$. Thus, from the purely observational
evidence we conclude that V440~Per does not pulsate in the
fundamental mode. Consequently, it must be an overtone Cepheid.

\begin{figure}
\includegraphics[width=0.47\textwidth, angle=0]{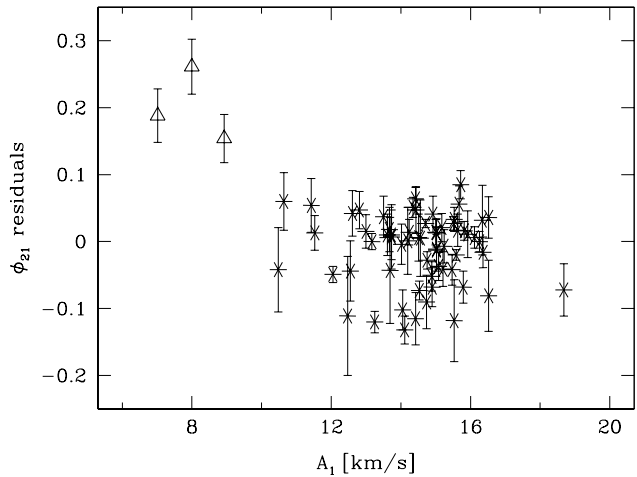}
\caption{Residuals from the mean fundamental mode $P-\phi_{21}$
         relation {\em vs.} pulsation amplitude $A_1$. Observational
         data points taken from \citet{kie99} and \citet{MGS09}.
         Same symbols as in Fig.\thinspace\ref{ASC3}.}
\label{ASC3a}
\end{figure}

The argument presented above critically depends on the assumption
that velocity $\phi_{21}$ (at a given period) does not depend on
the pulsation amplitude, which for V440~Per is lower than for the
fundamental mode Cepheids used as comparison. Such an assumption
is well justified by numerical computations ({\em i.e.} Buchler et
al. 1990; Smolec \& Moskalik 2008), nevertheless it should be
verified with available data. In Fig.\thinspace\ref{ASC3a} we plot
residuals from the mean fundamental mode $P-\phi_{21}$ relation
{\em vs.} amplitude $A_1$. In this case the mean relation is defined as
the parabola fitted to all fundamental mode Cepheids displayed in
Fig.\thinspace\ref{ASC3}. For $A_1$ in the range of
$10-17$\thinspace km/s there is no correlation of $\phi_{21}$
residuals with pulsation amplitude. For $A_1<10$\thinspace km/s
there is a weak evidence of $\phi_{21}$ increase. As the data are
very scarce, we are not convinced that this increase is
significant. If it were real, however, it would only strengthen our
conclusion that V440~Per strongly deviates from the fundamental
mode sequence.

\section{Modeling of V440~Per} \label{RSSEC1}

Classification of V440~Per as a first overtone pulsator, discussed
in the previous Section, is based solely on the morphological
properties of Cepheid velocity curves. This mode identification
can be strengthened by comparing the velocity curve of V440~Per with
those of hydrodynamical Cepheid models. Such a comparison was
first performed by \citet{kie99}, who used the unpublished radiative
models of Schaller \& Buchler (1994). They showed that the theoretical
progression of velocity $\phi_{21}$ supported the overtone
classification of V440~Per. However, their conclusion was somewhat
weakened by a large error of then available $\phi_{21}$ of this
star.

In the present section, we confirm and extend the results of
\citet{kie99}. It is easy to show, that the velocity curve of V440~Per
is incompatible with the fundamental mode Cepheid models. Indeed, all
published models display velocity $\phi_{21}$ higher that
3.0\thinspace rad {\em at all periods} and higher than
3.5\thinspace rad at the period of V440~Per ({\em i.e.} Buchler et
al. 1990; Moskalik et al. 1992; Smolec \& Moskalik 2008). This
holds true both for the convective models and for the older radiative
models. The computed velocity $\phi_{21}$ is very robust and shows no
sensitivity to the treatment of convection, choice of opacities or to
details of the numerical code. Most importantly, it is insensitive
to the pulsation amplitude ({\em cf.} Figs.\thinspace 8, 11 of
Smolec \& Moskalik 2008). Clearly, the only chance to match the
observed velocity curve of V440~Per is to search for an
appropriate overtone model.

With this goal in mind, we computed several sequences of the convective
overtone Cepheid models. We show that V440~Per fits theoretical
$\phi_{21}$ progression of the first overtone Cepheids and that
its velocity Fourier parameters can be accurately reproduced with
hydrodynamical computations.


\begin{table}
\begin{center}
\caption{Convective parameters of pulsation models discussed in
         this paper. $\alpha_s$, $\alpha_c$, $\alpha_d$, $\alpha_p$
         and $\gamma_r$ are given in the units of standard values
         (see Smolec \& Moskalik 2008). In the last column we give a
         linear upper limit for first overtone period
         ($P_{\mathrm{max}}$).
\label{RStab1}}
\begin{tabular}{cccccccccc}
\hline
Set & $\alpha$
          & $\alpha_m$
                & $\alpha_s$
                      & $\alpha_c$
                            & $\alpha_d$
                                  & $\alpha_p$
                                        & $\alpha_t$
                                               & $\gamma_r$
                                                     & $P_{\mathrm{max}}$ \\
\hline
A   & 1.5 & 0.1 & 1.0 & 1.0 & 1.0 & 0.0 & 0.00 & 0.0 & ~8.1$^\mathrm{d}$ \\
B   & 1.5 & 0.5 & 1.0 & 1.0 & 1.0 & 0.0 & 0.00 & 1.0 & ~8.6$^\mathrm{d}$ \\
C   & 1.5 & 0.1 & 1.0 & 1.0 & 1.0 & 1.0 & 0.01 & 0.0 & 13.5$^\mathrm{d}$ \\
\hline
\end{tabular}
\end{center}
\end{table}

Modeling of such a long period overtone pulsator is not an easy
task. The satisfactory models have to reproduce Fourier parameters and the
long period of this variable. The current hydrocodes used for modeling
of the radial pulsations adopt a time-dependent convection models (e.g.
Stellingwerf 1982, Kuhfu\ss{} 1986). These models introduce several
dimensionless parameters, that should be adjusted to match the
observational constraints. V440~Per with its exceptionally long
period, offers an opportunity to impose interesting limits on the
free parameters of the convection treatment.

In our computations we use the convective hydrocode of \citet{SM08}. The
code adopts \citet{K86} dynamical convection model reformulated for
use in stellar pulsation calculations. Kuhfu\ss{} model is
physically well motivated and self-consistent. It contains 8 scaling
parameters, summary of which, together with the detailed description of
the model equations is provided by \citet{SM08}. The values of parameters
used in the present paper are given in Table~\ref{RStab1}. All our
Cepheid models are constructed in the way described in \citet{SM08}.
In all computations we use Galactic chemical composition ($X=0.70$,
$Z=0.02$) and OPAL opacities \citep{IR96} computed with the
\citet{GN93} mixture. We use the mass-luminosity ($M\! -\! L$) relation
derived from \citet{Sch92} evolutionary tracks
($\log(L/L_\odot)=3.56\log(M/M_\odot)+0.79$).

\subsection{Pulsation period of V440~Per} \label{RSSEC11}

\begin{figure}
\includegraphics[width=84mm]{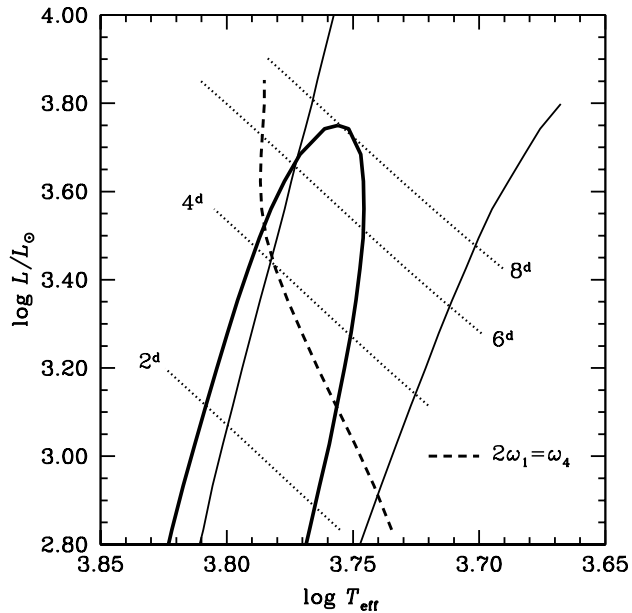}
\caption{Theoretical instability strips for convective
         parameters of set A.}
\label{RShr}
\end{figure}

Theoretical instability strips (IS) computed with the linear pulsation
code for set A of the convective parameters are presented in
Fig.\thinspace\ref{RShr}. Instability strips for the first overtone
and for the fundamental mode are enclosed with the thick and thin lines,
respectively. The dotted lines correspond to constant values of the
first overtone period, as indicated in the Figure.

First overtone IS does not extend toward an arbitrarily high
luminosity. The linear computations yield an upper limit for the first overtone pulsation
period, reaching $\sim
8.1$\thinspace day in Fig.\thinspace\ref{RShr} (see also
Table~\ref{RStab1}). This limit depends mostly on the adopted
convective parameters, but also on the $M\! -\! L$ relation and on
the metalicity. However, to find a model, that satisfies V440~Per
period constraint, it is not enough to assure appropriate linear
period limit ($P > 7.5$\thinspace day). This is because the maximum
overtone period at {\it full amplitude pulsation} is determined by the
nonlinear effects.

Between the linear blue edges of the first overtone and of the
fundamental mode and below their crossing point, the first overtone pulsation is the
only possibility. As one can see in
Fig.\thinspace\ref{RShr}, such a region is very narrow. A significant
part of the overtone IS lays inside the fundamental mode IS. For higher
luminosity and overtone periods longer than 6\thinspace days,
the instability strips of both modes entirely overlap. 

In the region,
where both modes are linearly unstable, the final pulsation
state is determined by the nonlinear effects (modal selection).
Full amplitude pulsation in one of the modes usually suffices to saturate the instability. 
In principle, a double-mode pulsation is also possible, but it
is very unlikely at long overtone periods that are of interest in
this paper (no double-mode pulsator with such a long overtone period
is known either in the Galaxy, or in the Magellanic Clouds; see
Soszy\'nski et al. 2008, Moskalik \& Ko\l{}aczkowski, 2008).
Therefore, construction of models matching period of V440~Per
requires both linear and nonlinear computations and can be used to
constrain parameters of convection model we use.

It is hard to constrain all 8 parameters entering the model. As was
shown by \citet{Y98}, different combinations of the convective $\alpha$
parameters may yield essentially the same results. Therefore, we
decided to freeze four parameters, for which standard values are in
use. In all studied sets of the convective parameters
(Table~\ref{RStab1}), mixing-length is set to $\alpha=1.5$ and for
$\alpha_s$, $\alpha_c$ and $\alpha_d$ we use the standard values (see
Smolec \& Moskalik 2008 for details). Set A represents the simplest
convection model, without turbulent pressure, turbulent flux or
radiative losses. These effects are turned on in set B (radiative
losses) and in set C (turbulent pressure and turbulent flux). One of
the crucial factors of the convection model is eddy-viscous damping,
the strength of which is determined by the $\alpha_m$ parameter. The lower
the eddy-viscous damping is (lower $\alpha_m$), the more linearly
unstable become the models and, consequently, the higher are their
pulsation amplitudes.

The most interesting outcome of the linear computations are the
period limits, $P_{\mathrm{max}}$, given in the last column of
Table~\ref{RStab1}. At the linear theory level, all sets of
convective parameters listed in the Table can satisfy V440~Per
period constraint.  This was assured by adjusting eddy-viscous
damping parameter, $\alpha_m$. In case of sets A and C (no radiative
losses) low values of $\alpha_m$ are required.

\begin{figure}
\includegraphics[width=0.47\textwidth]{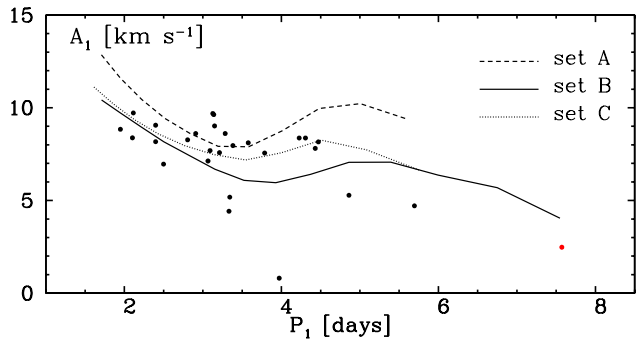}
\caption{Fourier amplitude $A_1$ for three sequences of overtone Cepheid
         models. Model amplitudes scaled by constant projection
         factor of 1.4. Except of V440~Per, observational data
         points are taken from \citet{kie99} and \citet{MGS09}.}
\label{RSperiodc}
\end{figure}

For each parameter set we computed a sequence of nonlinear models,
running parallel to the first overtone blue edge, at a constant
distance $\Delta T$. In this Section we use $\Delta T\! =\!
75$\thinspace K. This choice is arbitrary, but we believe, that such
models should reproduce most of the observed first overtone
variables. The static models were initialized in the first overtone mode
(as described in Smolec \& Moskalik 2008) and their time evolution was
followed, until the final pulsation state was reached. The computed
radial velocity curves were then decomposed into Fourier series. In
Fig.\thinspace\ref{RSperiodc} we plot amplitude $A_1$ {\em vs.}
pulsation period for three sequences of models. The most
satisfactory results are obtained for set B of the convective parameters. 
The pulsation amplitudes for this sequence are moderate and they match well
the observations (this conclusion will be strengthened in
Sect.\thinspace\ref{RSSEC12}). Also, for this sequence we obtain
the overtone models with the longest periods. Sequences computed with
the parameter sets A and C extend to periods not exceeding 6\thinspace
days. All more massive models ({\em i.e.} with longer overtone
periods) switched into the fundamental mode pulsation and therefore are
not plotted in the Figure. The amplitudes of the models computed with the
parameter sets A and C are larger. This is due to the adopted low value of
the eddy-viscous damping, needed to obtain high
enough limit for the linear overtone period. Any further decrease of $\alpha_m$
does not help. Although $P_{\mathrm{max}}$ increases, yet
models switch into the fundamental mode at periods well below of
7.57\thinspace days required to match V440~Per.  Additionally, at the short periods
the amplitudes of the first overtone models become
unacceptably high. We conclude, that inclusion of the radiative losses
is necessary to reproduce the long overtone period of V440~Per. As
we will demonstrate in Sect.\thinspace\ref{RSSEC12}, with the set B of the convective
parameters we can also reproduce the Fourier parameters of
V440~Per, as well as the overall progression of Fourier parameters for
all observed overtone Cepheids.

Our conclusion is consistent with the results of \citet{SZ07}. For
convection with the radiative losses\footnote{Note however, that in
Florida-Budapest hydrocode used by \citet{SZ07}, radiative losses
are modeled in a different way (see also Smolec \& Moskalik 2008).},
their models extend toward high overtone periods (parameter set A of
Szab\'o et al. 2007). Without radiative losses (set B of Szab\'o et
al. 2007) and with much higher eddy-viscous dissipation than in our
set A, their longest overtone periods fall below 3\thinspace
days. Comparing our models with the overtone Cepheid models of
\citet{FEU00} we note some inconsistency. \citet{FEU00} obtain long
overtone periods for all sets of convective parameters considered in
their paper. Their results are inconsistent with the results of
\citet{SZ07} (set A of Feuchtinger et al. 2000 and set B of Szab\'o
et al. 2007 are identical, but their linear results are
significantly different) as well as with our results. We trace this
discrepancy to different evaluation of the superadiabatic gradient,
$Y=\nabla-\nabla_a$, in our and in the Vienna code used by
\citet{FEU00}. In the Vienna code, the radiative pressure contribution
is neglected in computation of $\nabla_a$ \citep{FEU99}. This leads
to higher values of $\nabla_a$ ($\approx\! 0.4$ far from partial
ionization regions, see Figs. 3 and 4 of Wuchterl \& Feuchtinger
1998) and lower values of the superadiabatic gradient. Consequently, in
their models convection is less pronounced and the models are more
unstable. In fact, we were able to reproduce their linear results
when we neglected the contribution of radiation in the evaluation of
$\nabla_a$.

\begin{figure}
\includegraphics[width=0.47\textwidth]{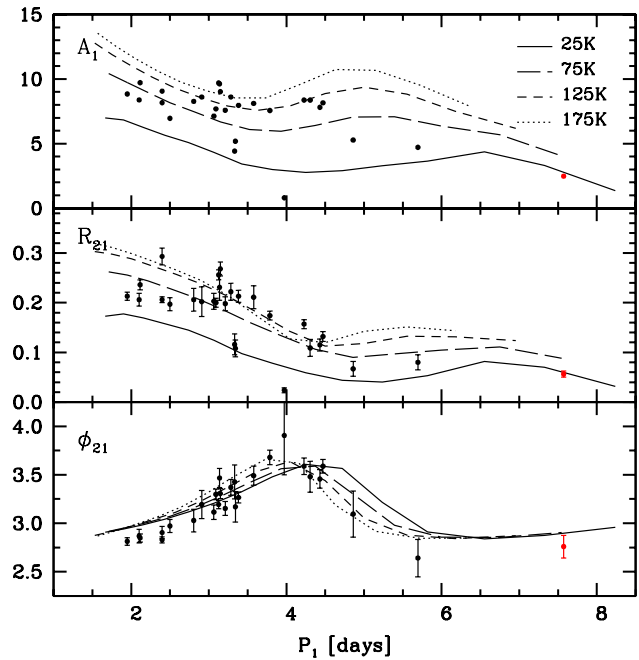}
\caption{Fourier parameters of Cepheid radial velocity curves. Model
         sequences are computed at constant distances from the first
         overtone blue edge, as indicated in the Figure. Except of
         V440~Per, observational data points are taken from
         \citet{kie99} and \citet{MGS09}. Error bars for most of the
         amplitudes are smaller than the symbol size and are not
         plotted.}
\label{RSfourier}
\end{figure}

\subsection{Fourier parameters of V440~Per}  \label{RSSEC12}

In this Section we focus our attention on set B of the convective parameters.
As demonstrated in the previous Section, this parameter set
allows to reproduce the long overtone periods and the observed pulsation
amplitudes. In Fig.\thinspace\ref{RSfourier} we compare the velocity
Fourier parameters of Galactic overtone Cepheids with those of
the hydrodynamical models computed with the parameters of set B. Four
sequences of models displayed in the plot run at different distances
from the first overtone blue edge ($\Delta T = 25$\thinspace K,
75\thinspace K, 125\thinspace K, 175\thinspace K). The overall
observed progression of Fourier parameters is well reproduced with our
models. For $\phi_{21}$, we notice that the model sequences are shifted
toward somewhat shorter periods. This can be easily explained. The
characteristic progression of velocity $\phi_{21}$ with period
is caused by the 2:1 resonance between the first overtone and the
fourth overtone \citep{kie99, FEU00}. The center of this resonance
is located at $P=4.2-4.6$\thinspace day. The exact location of the
resonance in hydrodynamic models depends mostly on the chosen
mass-luminosity relation. This relation was not adjusted in our
calculations to match the observed resonance
progression\footnote{Adjustment of $M\! -\! L$ relation to satisfy
the resonance constraint is not an easy task. Both the slope and the
zero point of the adopted $M\! -\! L$ relation affect location of
the resonance and the shape of the $\phi_{21}$ progression. In a
forthcoming publication (Smolec \& Moskalik, in preparation) we
address this problem in detail.}. As shown in
Fig.\thinspace\ref{RShr}, the resonance center (displayed with the
dashed line) crosses the center of the first overtone IS for periods
shorter than 4\thinspace days, which explains the horizontal shift in
Fig.\thinspace\ref{RSfourier}. Taking into account this shift,
amplitudes, amplitude ratios and Fourier phases agree satisfactorily
with observations, although at the short periods theoretical $\phi_{21}$
values are slightly too high. Considering V440~Per, we see the very good
agreement with the model sequence closest to the blue edge of the
overtone instability strip ($\Delta T\! =\! 25$\thinspace K). An
exact match can be easily obtained for the model sequence located
slightly closer to the blue edge. We note, that the progression of
the overtone $\phi_{21}$ at long periods is insensitive to the choice of
$\Delta T$. The models predict slightly higher $\phi_{21}$ than
actually observed in V440~Per, but the two values are consistent
within the error bar.

\section{Conclusions} \label{s4}

Using our new {\PST} we obtained 158 high precision radial velocity
observations of a low amplitude Cepheid V440~Per. We constructed
the pulsation velocity curve of V440~Per and we were able to reliably
detect its 1st harmonic, with amplitude of only $140\pm
15$\thinspace m/s. The measured Fourier phase $\phi_{21}=2.76\pm
0.12$\thinspace rad differs from the values observed in fundamental
mode Cepheids of a similar period by 7.25$\sigma$. Thus, we
demonstrated on purely morphological ground that V440~Per does not
pulsate in the fundamental mode. This settles the dispute between
\citet{SZ07} and \citet{kie99} and allows to classify V440~Per as an
overtone pulsator, the one with the longest period identified so
far ($P=7.57$\thinspace day).

Our results demonstrate that with suitable care, our inexpensive
instrument featuring MUSICOS Echelle spectrograph and a small
robotic telescope can achieve stability and precision surpassed only
in the extrasolar planet searches. Note, that we employed neither
the iodine cell nor environment control. Yet, our observations prove
that the secure mode identification is feasible even for the very low
amplitude Galactic Cepheids.

The overtone pulsation of V440~Per has interesting theoretical
consequences. To investigate them we employed our convective
linear and nonlinear pulsation codes \citep{SM08}. The first
overtone linear models are already constrained by the value of
the pulsation period at the center of the $2\omega_1=\omega_4$
resonance \citep{kie99, FEU00}. The very existence of V440~Per
imposes additional constraints. Namely, to linearly excite the
first overtone and then to obtain a {\em nonlinear full-amplitude}
overtone pulsation of such a long period, one has to fine tune the
dynamical convection model used in the pulsation calculations.
Our numerical experiments demonstrate that in the convective energy transport
radiative losses must be properly accounted for to maintain consistency with the observations.
With this effect taken into account,
the nonlinear overtone Cepheid models not only reproduce the
exceptionally long period of V440~Per, but they also reproduce neatly
all Fourier parameters of its pulsation velocity curve. No such
agreement can be achieved with models pulsating in the fundamental
mode. These results of the hydrodynamical modeling provide additional
support for our empirical classification of V440~Per as a first
overtone pulsator.

\section*{Acknowledgments}

This publication is based on the first results from {\PST}, a
large instrumental project developed at A.~Mickiewicz University
in Pozna\'n in the course of many years. Recent work by RB, MF and ASC 
was supported from the Polish MNiI grant N~N203~3020~35. We acknowledge
with gratitude permission to use facilities of- and help from- the
Borowiec Geodynamical Observatory of Centre for Space Research
(CBK PAN, Borowiec), in general, and encouragement from Prof.
Stanis{\l}aw Schillak, in particular. RS and PM work on the
pulsation codes is supported by the MNiSW Grant 1~P03D~011~30. We
would like to acknowledge inspiration and support by late Prof.
Bohdan Paczynski. Dr. Jacques Baudrand kindly supplied us with the
blueprints of MUSICOS.

\bsp
\appendix

\section{Fourier coefficients and their errors}\label{sa}

It seems useful to collect in one place all the relevant formulae for
Fourier coefficients of velocity $v(t)$ and their errors $\sigma$.
Defining Fourier series in the following way
\begin{eqnarray}
 v(t)-a_0 &=& \sum_{n=1}^{N}A_n\sin(n\omega t+\phi_n)\nonumber \\
~&=& \sum_{n=1}^{N}(a_n\cos{n\omega t}+b_n\sin{n\omega t})\label{ea.1}
\end{eqnarray}

\noindent where $N$ is the order of the fit, we obtain
\begin{equation}
A_n=\sqrt{a_n^2+b_n^2}\mbox{~~~and~~~}R_{n1} \equiv \frac{A_n}{A_1}\label{ea.2}
\end{equation}
\begin{equation}
\tan{\phi_n}=\frac{a_n}{b_n}\mbox{~~~and~~~}  \phi_{n1} \equiv \phi_n-n\phi_1\label{ea.3}
 \end{equation}

\noindent From variations of these equations we find

\begin{eqnarray}
 \delta A_n &=& \frac{a_n\delta a_n+b_n\delta b_n}{A_n}\label{ea.4}\\
 \delta R_{n1} &=& R_{n1}\left( \frac{a_n\delta a_n+b_n\delta b_n}{A_n^2}-
\frac{a_1\delta a_1+b_1\delta b_1}{A_1^2}\right)\label{ea.5}\\
 \delta \phi_n &=& \frac{b_n\delta a_n-a_n\delta b_n}{A_n^2}\label{ea.6}\\
 \delta \phi_{n1} &=& \frac{b_n\delta a_n-a_n\delta b_n}{A_n^2}-
n\frac{b_1\delta a_1-a_1\delta b_1}{A_1^2}\label{ea.7}
\end{eqnarray}

\noindent where RHS are scalar products of the vector $(\delta
a_n,\delta b_n,\delta a_1,\delta b_1)\equiv\delta F$ and the
corresponding derivatives $(\partial/\partial a_n,\partial/\partial
b_n,\partial/\partial a_1,\partial/\partial b_1)\equiv L[\cdot]$.
For example $\delta \phi_{n1}=L[\phi_{n1}]\cdot\delta F$ and third
component of vector $L[\phi_{n1}]$ is $\partial\phi_{n1}/\partial
a_1=-nb_1/A_1^2$. By the law of error propagation, variance of
$\phi_{n1}$ is

\begin{equation}
\sigma^2_{\phi_{n1}}\equiv Var[\phi_{n1}]=L[\phi_{n1}]\times Cov[a_n,b_n,a_1,b_1]\times L[\phi_{n1}]^T\label{ea.8}
\end{equation}

\noindent where $Cov[\cdot]$ denotes the covariance matrix of raw
Fourier coefficients (Eq.\thinspace\ref{ea.1}), $\times$ denotes
matrix multiplication and superscript $T$ indicates matrix
transpose. Analogous expressions hold for other coefficients. The
complete linearized transformation matrix $M$ is obtained by
substitution of the $L$ vectors corresponding to Eqs. (\ref{ea.4}--\ref{ea.7}) 
as its rows. The full covariance matrix of all Fourier coefficients is 
seldom needed in practice. It may be obtained from Eq. (\ref{ea.8}) 
by substitution of $M$ for $L$.

\label{lastpage}
\end{document}